# 502 Gbits/s quantum random number generation with simple and compact structure

JINLU LIU[1], JIE YANG[1], ZHENGYU LI[2], WEI HUANG[1], BINGJIE XU[1*], HONG GUO[2†]

[1]Science and Technology on Security communication Laboratory, Institute of Southwestern Communication, Chengdu 610041, China
[2]State Key Laboratory of Advance Optical Communication Systems and Networks, Center for Computational Science & Engineering (CCSE) and Center for Quantum Information Technology, School of Electronics Engineering and Computer Science, Peking University, Beijing 100871, China
*Corresponding author: xbjpku@pku.edu.cn

†Corresponding author: hongguo@pku.edu.cn

**We propose and implement a simple and compact quantum random number generation (QRNG) scheme based on the quantum phase fluctuations of a DFB laser. The distribution probability of the experimentally measured data fits well with the simulation result, got from the theoretical model we established. Min-entropy estimation and Toeplitz-hashing randomness extractor are used to obtain the final random bit. The proposed approach has advantages not only in simple structure but also in high random bit generation rate. As a result, 502 Gbits/s random bits generation speed can be obtained, which is much higher than previous similar schemes. This approach offers a possibility to promote the practical application of QRNG.**

Random numbers play a key role in many fields of science and technology, such as statistical sampling, randomized algorithm and cryptography [1]. According to different generation methods, random numbers can be divided into pseudo-random numbers (PRN) and physical random numbers. PRN are generated by complex algorithms, which is almost impossible to decode in practical terms in the past. However, with the rapid development of quantum computation, PRN may leave a loophole in application, especially in the field of information security [2]. Different from PRN, physical random numbers are generated by measuring non-deterministic physical process, which is also called true random numbers. Over the past decades, quantum random number generation (QRNG), as the main method with provable security to generate true random numbers, is widely studied [3-16]. Based on different physical model, many approaches of QRNG have been proposed. Measuring reflection or transmission of a single photon with single photon detection (SPD), was adapted in the early stage of QRNG [3, 4, 5], where the random bit generation rate (RBGR) was extremely limited by the count rate of the SPD [13]. By measuring the vacuum fluctuation with shot-noise limited homodyne detection, a 2G bps real-time QRNG has been demonstrated

[9], where the RBGR has been greatly improved. Nonetheless, it still cannot satisfy the requirement of high RBGR application.

Another method is applying traditional high bandwidth photo-detector to measure the phase fluctuation or amplified spontaneous emission (ASE) noise [6, 7, 11, 12, 15, 17], to obtain higher RBGR. As a random source, laser phase fluctuation originates from spontaneous emission which is a non-predictable quantum-mechanical process. In the QRNG scheme based on the phase fluctuation, an interferometer is commonly used to convert phase fluctuation into intensity fluctuation. By using a classical photo-detector and high speed analog-to-digital converter (ADC), a relatively high RBGR can be obtained. For instance, a 68G bits/s QRNG by measuring laser phase fluctuation has been reported [12], where two faraday rotator mirrors were used to remove the polarization effects, and a phase shifter with complex feedback control was needed to keep the interference stable.

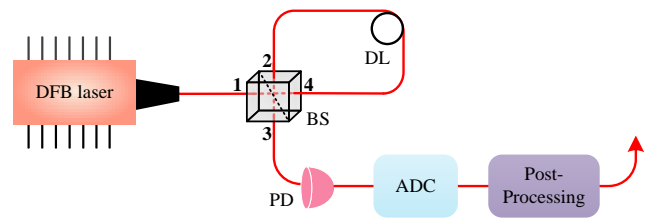

Fig.1. Experimental setup of the compact QRNG. A DFB laser diode emits CW beam. It enters one input port (port 1) of a BS, then part of the beam is directly coupled into the PD, and the other part is coupled into a fiber loop, which is composed by the BS and DL, and recirculates inside the loop. Each pass of the loop introduces a time delay by the DL, and the beam from port4 will interfere with the beam from the LD. An ADC is used to extract random bits from PD. Employing post-processing, the final random bits can be obtained. CW: continuous wave, BS: 2×2 50/50 polarization-maintaining beam splitter, DL: delay line, PD: photo-detector with high bandwidth, ADC: 8-bit analog-to-digital converter.

In this paper, we report a 502 Gbits/s QRNG scheme with simple and compact structure based on measuring quantum phase fluctuation. Firstly, this scheme only employs one beam splitter with a delay line to accomplish interference, where the complexity is significantly reduced compared with previous similar schemes. Secondly, the experimental results show that the scheme needs not complex feedback control to keep the interference stable. Furthermore, a theoretical model of the system is established, and the simulation results of the probability distribution fit well with the experimental data. After post-processing, the final random bits can pass the standard randomness tests.

Experimental setup of the proposed QRNG is shown in Fig.1. A DFB laser diode (LD), center wavelength at 1550.12nm, driven by a butterfly packaged laser diode driver, emits continuous-wave (CW) beam. To maximize the quantum phase fluctuation, the LD is operated around its threshold level. The CW beam is split into two paths by a 2×2 50/50 polarization-maintaining beam splitter (BS). One beam is directly coupled into a 40GHz photo-detector (PD, U²T, XPDV2120RA). The other beam is coupled into the fiber loop, composed by the BS and a 4m delay line (DL), and recirculates inside the loop. Each pass of the loop introduces a 20ns time delay by the 4m DL, and the beam after $N$ circulations has $20N$ ns delay time. Employing the build-in-8-bit ADC in a high speed oscilloscope (Agilent, DSAV334A, 33GHz bandwidth) to sample the output of PD, raw random bits are generated. Applying min-entropy estimation and Toeplitz randomness extractor [11] to post-process the raw data, the final random bits can be obtained.

Theoretical model of the setup is established as follows. The field at port 1 of the BS is

$$E_{p1}(t) = A\exp[i\omega t + i\varphi(t)] \quad (1)$$

where A is the amplitude of electric field, $\omega$ is the optical center angular frequency, and $\varphi(t)$ is the phase fluctuation of the laser, respectively. The field at port 2 of the BS is the sum of beams that have transmitted in the delay loop for different times of circulations ranging from 1 to N (N → ∞), which includes all the components with significant values and can be calculated as

$$E_{p2}(t) = A\sum_{k=1}^{N}\left(\frac{1}{\sqrt{2}}\right)^k \exp[i\omega(t-k\Delta t) + i\varphi(t-k\Delta t)] \quad (2)$$

where $k \in [1, N]$ is an integer, and $\Delta t$ is the time delay induced by DL. $E_{p1}$ and $E_{p2}$ will interfere at BS and the optical intensity, detected by the PD, can be given by

$$I = 1 + \sum_{k=1}^{n}\left(\frac{1}{2}\right)^k$$
$$+ \sum_{k=1}^{N}(\sqrt{2})^{2-k}\left(-\cos(k\omega\Delta t + \Delta\varphi_N^k)\right)$$
$$+ \sum_{k=1}^{N}(\sqrt{2})^{2-k}\sum_{j=1}^{k-1}\left(\frac{1}{\sqrt{2}}\right)^j \cos[(k-j)\omega\Delta t + \Delta\varphi_{N-j}^{k-j}] \quad (3)$$

where $j$ and $k$ are integers, and $k \in [1, N]$, $j \in [1, k)$, $\Delta\varphi_N^k = \varphi(N\Delta t) - \varphi((N-k)\Delta t)$ is the phase fluctuation between $N$-th order circulation and $N$-$k$-th circulation laser beam. Note that the DC component is $1 + \sum_{k=1}^{n}(1/2)^k$, which has no contribution to the generation of random bits. $\Delta\varphi_N^k$ is a Guassian random variable due to spontaneous emission[18, 19], therefore $I$, which is a superposition of $\Delta\varphi_N^k$, can be quantified to random bits.

Based on the above theoretical model, the measured intensity $I$ is simulated. In the setup, the attenuation of the optical loop is 0.3dB, and the time delay $\Delta t = 20ns$, the line width of the laser working around threshold level is $5.5MHz$, and $\Delta\varphi$ is modeled as $\Delta\varphi \sim N(0, 2\pi\Delta f \Delta t)$. The simulation result (red line) is shown in Fig.2., which is well limited in the range of confidence intervals (blue error-bar) evaluated by experimental measured data.

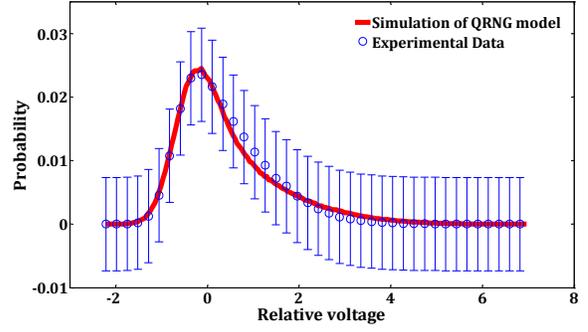

Fig.2 The normalized distribution of experimentally measured interference intensity and the simulation result of the theoretical model for the proposed QRNG.

Furthermore, long time scale experimental results show that the proposed scheme doesn't need complex feedback control to keep the interference stable and randomness extraction rates. Firstly, all fibers used in the QRNG system is polarization maintaining, where polarization change can be ignored. Secondly, we monitor the interference output from PD for 3 hours, and the statistical distribution is shown in Fig.3, where the probability distribution and the minimal entropy of the measured data are relatively stable. Therefore, the quantum phase fluctuation is retained and slow environmental fluctuation does not destroy the stability of the system.

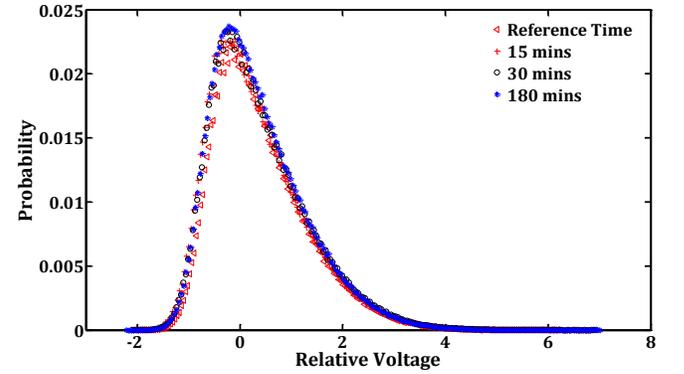

Fig.3 The distribution of experimental measured data at different time.

After sampling by ADC, we apply min-entropy evaluation on the raw data. Min-entropy quantifies the amount of the randomness of a distribution in worst-case scenario, which is defined as

$$H_\infty(X) = -\log_2(\max_{x\in\{0,1\}^n} \Pr[X=x]) \quad (4)$$

In experiment, the min-entropy of the raw data is $H_\infty(X) = 6.27$ per Byte. Therefore, 6.27 random bits can be extracted from each sample. A fast Fourier transform (FFT) based Toeplitz-matrix hash function with a size of $n \times m$ ($n = 5\times 10^6$ and $m = 3.92\times 10^6$) is applied to extract randomness from raw random bits. By employing a high bandwidth PD and a bulid-in-8-bit ADC of 80G sampling rate, the final RBGR reach highly up to 502Gbits/s.

In order to evaluate the randomness, we applied three randomness test batteries, NIST-STS, Diehard and ENT, and the corresponding test results are shown in Fig.4, Fig5 and table 1. The random bits generated by the proposed QRNG can pass the all tests.

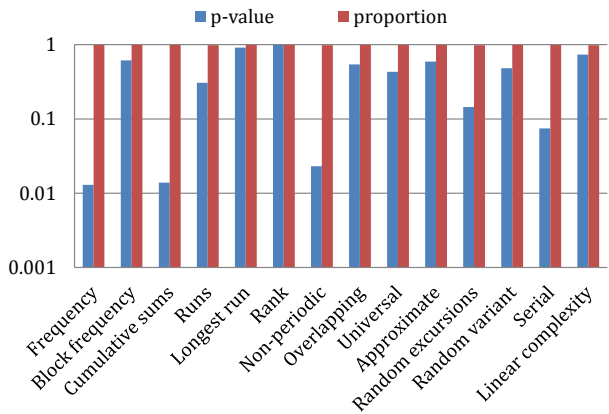

Fig.4. Results of the NIST-STS test suite for a 1Gbit sequence. The significance was set by 0.01. To past the test, p-value needs to satisfy $0.01 \leq p-value \leq 0.99$.

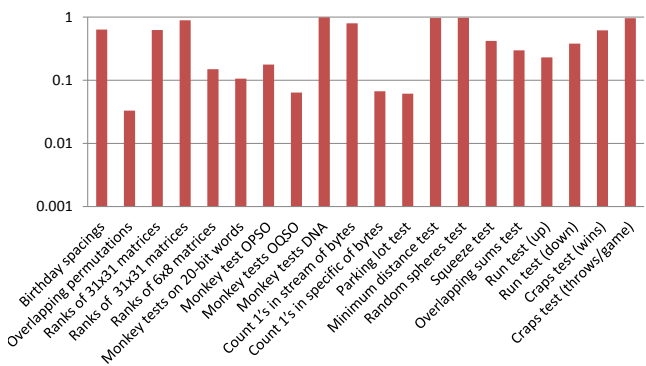

Fig.5. Results of the Diehard statistical suite for a 1Gbit sequence. For the case of multiple p-values in Diehard test suite, a Kolmogorov-Smirnov (KS) test is used to obtain a final p-value, which measures the uniformity of the multiple p-value.

Table1. Results of the ENT statistical suite for a 1Gbit sequence. ENT test evaluates the randomness of the random number sequence on macroscopic view, which represents the probability that the tested random number sequence is true randomness.

| |
| --- |
| Entropy = 1.000000 bit per bit |
| (the optimum compression would reduce the bit file by 0%) |
| $\chi^2$ distribution is 1.84 |
| (randomly would exceed this value by 17.50% of the times) |
| Arithmetic mean value of data bits is 0.5000 |
| (0.5 = random) |
| Monte Carlo value for $\pi$ is 3.141648974 |
| (error 0.00%) |
| Serial correlation coefficient is 0.000010 |
| ( totally uncorrelated = 0.0) |

To verify the stability of the QRNG in randomness, we test the experimental measured data which is sampled at different time. The data can pass all the randomness tests, which shows that the randomness is uninfluenced by the environmental fluctuation in a relatively long time.

In summary, we propose and implement a simple and compact quantum random number generation (QRNG) scheme based on the quantum phase fluctuations of a DFB laser. The distribution probability of the experimentally measured data fits well with the simulation result, got from the theoretical model we established. Min-entropy estimation and Toeplitz-hashing randomness extractor are used to obtain the final random bit. The proposed approach has advantages not only in simple structure but also in high random bit generation rate. As a result, 502 Gbits/s random bits generation speed can be obtained, which is much higher than similar schemes. Combined with the high speed post-processing technology [17], the proposed QRNG scheme has broad prospects, especially in high-speed and integrated application system.

**Funding.** National Science Foundation of China (NSFC) (61501414);